\documentclass{article}
\usepackage[T1]{fontenc} 
\usepackage[utf8]{inputenc} 
\usepackage{ismir,amsmath,cite,url}
\usepackage{graphicx}
\usepackage{color}
\usepackage[hang]{footmisc}

\usepackage{lineno}

\usepackage{multirow}
\usepackage[ruled, vlined]{algorithm2e}
\usepackage{subfigure}
\usepackage{amssymb}
\usepackage{mathtools, nccmath}
\usepackage{enumitem}
\usepackage{multicol}
\usepackage{booktabs}

\DeclarePairedDelimiterX{\infdivx}[2]{(}{)}{%
  #1\;\delimsize\|\;#2%
}
\newcommand{\infdiv}{\infdivx}

\def\authorname{Wayne Chi, Prachi Kumar, Suri Yaddanapudi, Rahul Suresh, Umut Isik}

\begin{document}

\title{Generating Music with a Self-Correcting Non-Chronological Autoregressive Model}

\multauthor
{Wayne Chi${\,}^{*1}$ \hspace{1cm} Prachi Kumar${\,}^{*1}$ \hspace{1cm} Suri Yaddanapudi$^1$} { \bfseries{Rahul Suresh$^1$ \hspace{1cm} Umut Isik$^1$ \hspace{1cm}}\\
  $^{*}$ Equal Contribution \hspace{1cm} $^1$ Amazon Web Services\\
{\tt\small waynchi@amazon.com, kumprach@amazon.com, yaddas@amazon.com}\\ {\tt\small surerahu@amazon.com,  umutisik@amazon.com}
}

\maketitle

\begin{abstract}

We describe a novel approach for generating music using a self-correcting, non-chronological, autoregressive model.
We represent music as a sequence of edit events, each of which denotes either the addition or removal of a note---even a note previously generated by the model.
During inference, we generate one edit event at a time using direct ancestral sampling.
Our approach allows the model to fix previous mistakes such as incorrectly sampled notes and prevent accumulation of errors which autoregressive models are prone to have.
Another benefit is a finer, note-by-note control during human and AI collaborative composition.
We show through quantitative metrics and human survey evaluation that our approach generates better results than orderless NADE and Gibbs sampling approaches.

\end{abstract}

\section{Introduction}\label{sec:introduction}

There have been two primary approaches to generating music with deep neural network-based generative models.
In the first class, music generation is essentially treated as an image generation problem 
\cite{dong2018musegan,yang2017midinet}.
In the second class, music generation is treated as a musical time series generation problem, analogous to autoregressive language modeling
\cite{huang2018music, child2019generating, dai2019transformer, payne2019musenet, wu2019hierarchical}.
The human process of music composition, however, is often non-chronological. 
Notes can be filled in anytime throughout the music piece to create new chords and melodies, add harmony, or embellish existing motifs.

In this work, we propose ES-Net\footnote{Code: https://git.io/esnet\\
Samples: https://git.io/esnet-samples
},
a method that uses elements from both the image-based and time series generation techniques.
Our method operates on piano roll images with a 2D convolutional neural network, but autoregressively adds or removes notes one at a time in an arbitrary, non-chronological order.
We model the conditional distribution of note add or remove events given pre-existing notes.
After sampling from the distribution, we re-input the resulting piano roll into the model to get the distribution of the next add and remove events. 
From a probabilistic point of view, this corresponds to considering each piano roll as obtained from a randomly ordered sequence of add and remove events and autoregressively modeling the distribution of such sequences of events. 

Poor samples due to accumulation of errors is a well-documented problem with autoregressive models\cite{bengio2015scheduled, huszar2015not, lamb2016professor, venkatraman2015improving},
especially when directly sampling from the conditional distribution (i.e. direct ancestral sampling).
While other sampling techniques such as Gibbs sampling \cite{huang2019counterpoint} can be used to bypass this problem,
we show that direct ancestral sampling is sufficient if the data representation includes removal of past samples.
This allows the model to detect previous mistakes and fix them.

Our primary use case is melody assistance for users generating musical compositions.
Users can feed in a melody as a conditional input and have the model generate musical accompaniments as well as fix any off-beat or out of tune inputs.
One distinct advantage of our approach is that it allows note-by-note control for users. 
A user can undo and redo the generation of individual notes 
or explicitly add and remove individual notes to collaborate with the model and guide the music composition process. 
Thus, this approach allows users to have a finer degree of control during sampling and better promotes human and AI collaboration.

The remainder of the paper is organized as follows.
In Section \ref{sec:related-works} we discuss related works.
In Section \ref{sec:problem-definition} we show how to model a distribution of musical pieces using a new representation of music.
In Section \ref{sec:training} we discuss our training procedure.
In Section \ref{sec:inference} we discuss our sampling procedure.
In Section \ref{sec:results} we provide empirical results in the form of quantitative metrics and human evaluation compared against other approaches.
Finally, in Section \ref{sec:future-work} and \ref{sec:conclusion} we describe future work and conclusions.

\section{Related Work}\label{sec:related-works}

Following the introduction of NADE \cite{larochelle2011neural, uria2016neural} and orderless NADE \cite{uria2014deep}
there have been several works built upon the concept of ordered and unordered autoregressive models.
Coconet---the algorithm behind Google's Bach Doodle\footnote{https://magenta.tensorflow.org/coconet}---is a machine learning model that also uses a convolutional model to generate music by adding counterpoints to existing user input \cite{huang2019counterpoint}. 
The difference with this work is that Coconet's inference uses Gibbs sampling rather than direct ancestral sampling. DeepBach \cite{hadjeres1612deepbach} generates Bach style chorales using pseudo-Gibbs sampling.
PixelCNN \cite{oord2016pixel} models an image autoregressively and generates pixels one by one in a pre-specified order while our generation is unordered.
In a NLP setting, recent works also explore non left-to-right ordering \cite{stern2019insertion, chan2019kermit} and deletion \cite{gu2019levenshtein}.

In general, there is a rich history of using deep learning to generate music\cite{briot2017deep}.
Many of them use autoregressive based approaches.
RNN-RBM models temporal dependencies to generate polyphonic music in a single track\cite{boulanger2013high}.
Hierararchical RNNs have been used to encode different features of pop music\cite{chu2016song}.
LSTMs were able to successfully model and generate music as well \cite{sturm2016music}.
Music Transformer is able to capture and generate music with long term structure and motifs \cite{huang2018music}.
So far, these approaches have been mostly chronological while ours is non-chronological.
While GAN-based approaches clearly differ from ours, these methods have shown the ability to generate high quality music. 
MuseGAN is a GAN-based approach for multi-track piano roll generation\cite{dong2018musegan}.
MidiNet uses a CNN-based GAN to generate music\cite{yang2017midinet}.
C-RNN-GAN generates music using a RNN based architecture with adversarial training \cite{mogren2016c}.
SeqGAN use GANs for sequence generation and apply it to music generation \cite{yu2017seqgan} .


\section{Problem Definition}\label{sec:problem-definition}

We consider a musical piece $x \in X$ as a point in $\{0,1\}^{T \times P}$ where 
$T$ is the number of time steps and $P$ is the number of note pitches.
This represents a simplified piano roll (PR)---a discrete representation of music as an image matrix across pitch and time.
There exists a probability density function $p^{\operatorname{PR}}(x)$ on $\{0,1\}^{T \times P}$ of musical pieces.
Note in particular that this does not model velocity and that 
notes adjacent in time are treated as one continuous held note;
we discuss ways to represent velocity and repeated notes in Section \ref{sec:future-work}.
Instead of modeling $p^{\operatorname{PR}}(x)$ on $\{0,1\}^{T \times P}$ directly, 
we model the distributions as $p^{\operatorname{ES}}(s)$ on the set of \textbf{edit sequences} (ES).
An \textbf{edit sequence} of length $M$ is a tuple of $M$-many \textbf{edit events} where an \textbf{edit event} is a matrix $e^{(t,p)} \in \{ 0,1\}^{T \times P}$ that has one entry equal to one, and all other entries equal to zero (i.e. a one-hot matrix).
We denote the set of all edit events by $\mathcal{E}$ and of edit sequences of length $M$ by $\mathcal{E}^M$.
The following maps edit sequences to piano rolls:
\begin{align}
    &\pi :  \bigcup\limits_{M=1}^{\infty} \mathcal{E}^{M} \rightarrow \{0,1\}^{T \times P} \label{eq:1} \\
    &\pi(e_{1}, \ldots, e_{M}) = \sum\limits_{i=1}^M e_i \pmod 2. \label{eq:2}
\end{align}
where \eqref{eq:2} allows edit events to handle either note addition or removal depending on if a previous edit event exists at the same time and pitch.

\begin{figure}[h]
    \centering
    \includegraphics[scale=0.4]{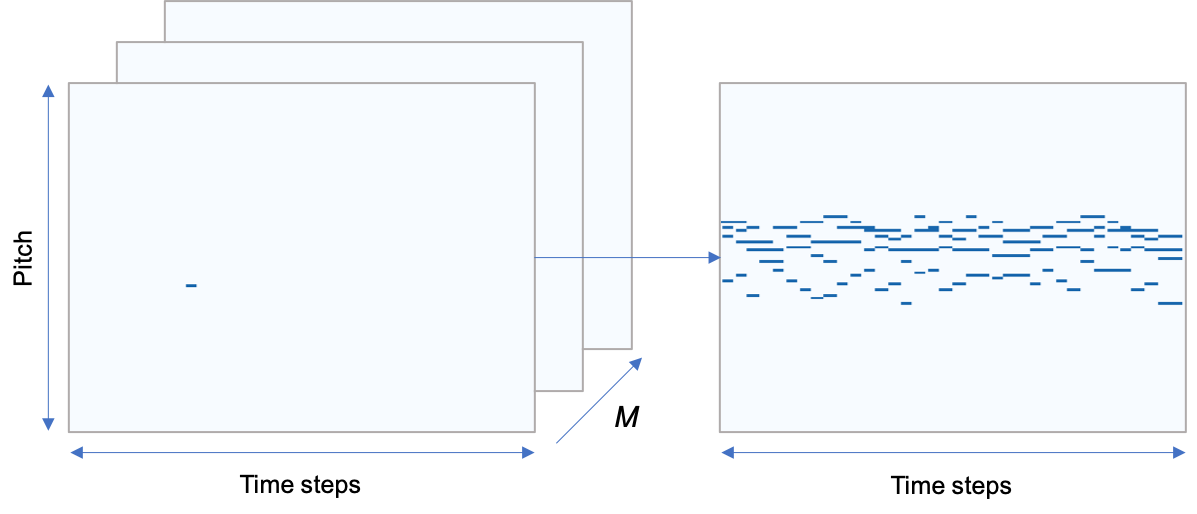}
    \caption{Mapping from an edit sequence (left) of length M to a piano roll (right). Each slice in an edit sequence is the addition or removal of a note.}
    \label{fig:mapping}
\end{figure}

The mapping between the two joint probability distributions is as follows:
\begin{equation}
\begin{aligned}
    p^{\operatorname{PR}}(x) 
    &= \ p^{\operatorname{PR}}(\{(t_1, p_1), \dots, (t_N, p_N)\}) 
    \\
    &= \ \sum\limits_{s \in \pi^{-1}(x)}^{\infty} p^{\operatorname{ES}}(s) 
\end{aligned}
\label{eq:mapping-joint}
\end{equation}
where $N$ is the number of notes in the piano roll, 
$(t_i, p_i)$ is the time and pitch of a note or edit event,
$\pi^{-1}(x)$ is the inverse image set of $\pi(x)$,
and $s$ is a sequence of edit events $(t_1, p_1) \ldots (t_M, p_M)$ where $M \geq N$.
We can further factorize $p^{\operatorname{ES}}(s)$ as:
\begin{equation}
\begin{aligned}
p^{\operatorname{ES}}(s) 
&= p^{\operatorname{ES}}\big((t_1, p_1), \ldots, (t_M, p_M)\big)\\
&= \prod_{i=1}^M
p^{\operatorname{ES}}\big((t_{i}, p_{i}) \vert 
(t_{1}, p_{1}), \ldots, 
(t_{i-1}, p_{i-1})\big) 
\end{aligned}
\label{eq:factorized}
\end{equation}
We assume that
$p^{\operatorname{ES}}((t_{i}, p_{i}) \vert  (t_{1}, p_{1}), \ldots, (t_{i-1}, p_{i-1}))$
is ordering invariant (i.e. the ordering of edit events in an edit sequence does not affect the resulting piano roll).

Our goal is to train a model to map the distribution of edit sequences $p^{\operatorname{ES}}(s)$.
By sampling autoregressively from $p^{\operatorname{ES}}(s)$,
we will generate a sequence of edit events that can be mapped back into a piano roll representation and then converted to MIDI.

\subsection{Orderless NADE} \label{sec:orderless-nade}

We compare our approach to orderless NADE which generates music by randomly choosing an ordering and sampling notes one by one until termination.
We can represent the iterative notewise addition of orderless NADE as a special case of edit sequences where edit events can only represent note addition. 
Let us call this distribution $p^{\operatorname{O-NADE}}(x)$.
Since notes are only added, $M = N$ for unconditioned generation; 
thus, there is a finite set of orderings and we can factorize $p^{\operatorname{O-NADE}}(x)$ as:
\begin{align*}
        \sum_{\sigma \in S_{N}}
        \prod_{i=1}^N
        p^{\operatorname{O-NADE}}\big(
        &(t_{\sigma(i)}, p_{\sigma(i)}) \vert \\
        &(t_{\sigma(1)}, p_{\sigma(1)}), ..., (t_{\sigma(i-1)}, p_{\sigma(i-1)})\big)
\end{align*}
where $S_N$ is the set of all permutations $\{1,2,...,N\} \rightarrow \{1,2,...,N\}$.
This factorization is equivalent to orderless NADE \cite{uria2014deep}.
In practice, the orderless NADE approach leads to poorer musical samples due to accumulation of errors which we confirm in Section \ref{sec:results}.


\section{Training}\label{sec:training}

\begin{figure*}[t]
    \centering
    \includegraphics[scale=0.3]{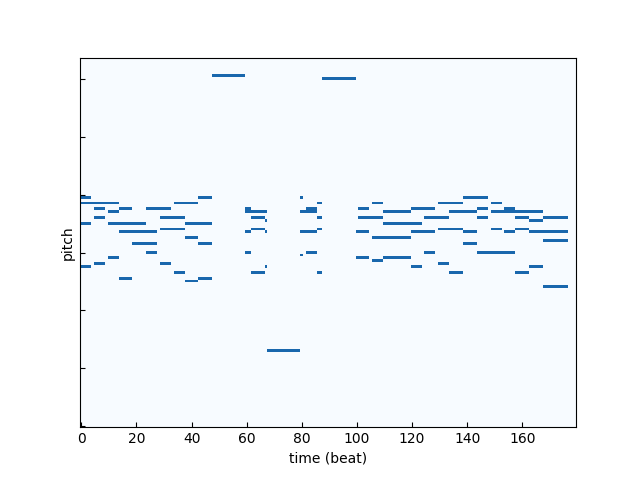}
    \includegraphics[scale=0.3]{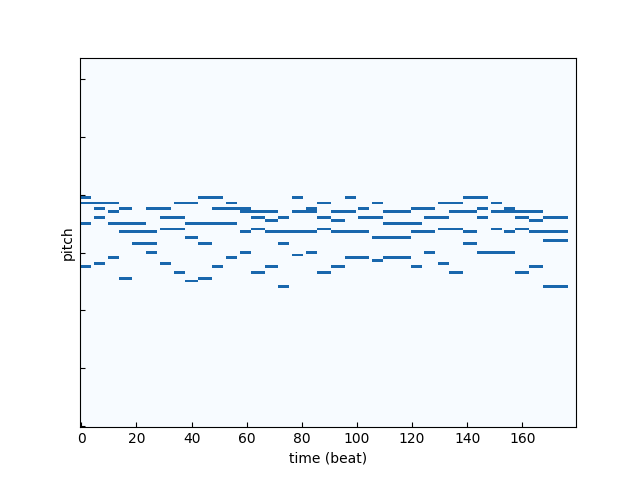}
    \includegraphics[scale=0.3]{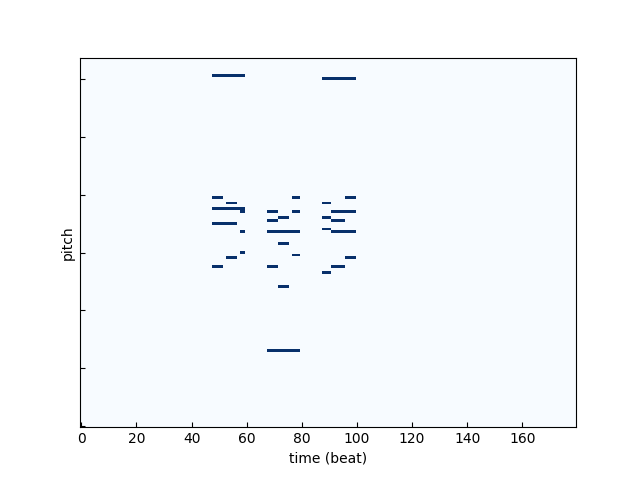}
    \caption{Input piano roll (left), target piano roll (middle), and symmetric difference between the input and target piano rolls (right)}
    \label{fig:symmetric-difference}
\end{figure*}

Given an input piano roll, $\mathcal{I}$, and target piano roll, $\mathcal{T}$,
we train our model to output the conditional probabilities
$p^{\operatorname{ES}}((t_{i}, p_{i}) \vert  (t_{1}, p_{1}), \ldots, (t_{i-1}, p_{i-1}))$ of the next edit event in edit sequences that can recreate the target from the input piano roll.
For each piano roll in the training set, we generate the input piano roll by 
(a) masking existing random notes and 
(b) adding extraneous random notes to
the target piano roll.
We train the model to recreate $\mathcal{T}$ from $\mathcal{I}$; 
augmentation (a) trains the model to add notes and
(b) trains the model to remove notes.
For each target piano roll, we generate multiple augmented inputs with varying number of notes masked and added, in order to train the model to handle varying number of differences between the input and target.
We find that masking between 0 to 100 percent of all existing notes and adding 0 to 1.5 percent of all possible extraneous notes gives us the best results.

Our goal is to have the model output the conditional probabilities for the next edit event. 
Since we assume ordering invariance in \eqref{eq:factorized},
we can also assume that every note difference between $\mathcal{I}$ and $\mathcal{T}$---whether it requires the addition or removal of a note---is equally likely to be the next edit event.
Thus, we model the distribution of edit events for the next step as the uniform distribution $\textit{U}$ supported on the symmetric difference $\mathcal{I} \Delta \mathcal{T}$ between $\mathcal{I}$ and $\mathcal{T}$ 
(i.e the exclusive or of each note between $\mathcal{I}$ and $\mathcal{T}$). 

We use the Kullback-Liebler divergence between $\textit{U}$ and the model's output distribution as the loss function:
\begin{equation} 
\mathcal{L}(\mathcal{I}, \mathcal{T}, P) = D_{KL}\infdiv{P}{U},
\label{theloss}
\end{equation}
where $P$ is the softmax over the model's logits at each time and pitch.
Normally, binary cross-entropy loss---where the label is the next note---would be used, but since we assume ordering invariance in \eqref{eq:factorized}, the next note is equally likely to be any of the future notes. 
Therefore, training with \eqref{theloss} is equivalent to training many times where the label is randomly chosen from future notes.

\subsection{Model}
We train a model based on the U-Net architecture \cite{ronneberger2015u}. 
This choice is not critical as our approach should generalize to other CNN architectures.
We describe our approach for reproducibility.
Our U-Net contains five downsampling blocks and five upsampling blocks.
In each block there contains a batch normalization layer, two 2D convolutional layers each with a 3x3 kernel,
a max pooling layer,
and a drop out layer with a 0.5 dropout rate.
We begin with 32 filters.
We double the number of filters after each downsampling block and halve the number of filters after each upsampling block.
We use the Adam optimizer \cite{kingma2014adam} with a learning rate of 0.001.
We use RELU for our activation function, except for the final layer where we output a linear activation at each time and pitch.
Finally, we apply softmax over the logits when calculating the loss and during sampling.

\section{Inference}\label{sec:inference}
We sample from the model’s output probabilities through direct ancestral sampling. We feed the input melody to the model, sample from the softmax over all times and pitches to determine the next edit event, modify the input melody based on that edit event, and then feed that melody back into the model.
We repeat this over multiple iterations and condition each time on our previous predictions. 
Since we do not differentiate between adding and removing notes during training, the sampling process is the same for any type of edit event.
We allow users to restrict the number of notes to remove;
this prevents the model from completely overwriting the original input.
We also allow users to control how many sampling iterations are performed.
Lastly, we allow the user to change the temperature during sampling. 
By changing the shape of the distribution, users can make compositions more or less ``creative'' at the risk of lowering quality.
We surface these hyperparameters to allow users to more freedom and customizability when generating music compositions.

\section{Empirical Evaluation}\label{sec:results}
We compare our approach against orderless NADE and Gibbs sampling using quantitative metrics and human survey evaluation. 
We also describe a notewise approximate log likelihood calculation for our approach and explain why log likelihood is not a good metric for comparing our approach to orderless NADE.
We build an orderless NADE model using the approach described in Section \ref{sec:orderless-nade} and training with only masked notes.
We use Coconet \cite{huang2019counterpoint} to represent Gibbs sampling.
While our main focus is to only use Coconet for sampling technique comparisons, there are a few notable differences between Coconet and our approach.
First, Coconet does not explicitly train the model to remove notes, but notes---including the input---may be removed during the Gibbs sampling masking process; 
our approach explicitly models note removal.
Second, Coconet assumes that there are four instruments and that ``each instrument plays exactly one pitch a time'' \cite{huang2019counterpoint};
our approach has no such constraint and can generate music across all times and pitches.
Third, Coconet trains a CNN that preserves the same size for each layer;
we train a model based on the U-Net architecture.
Since Coconet is trained on the JSB Chorales dataset, we evaluate our results and Orderless NADE's results using the same dataset and the same train-val-test split in order to provide a fair comparison.
For all other parameters (e.g. temperature), we maintain identical settings for each approach in order to benchmark fairly.


\subsection{Data}\label{sec:data} 
We use the Infinite Bach dataset\footnote{https://github.com/jamesrobertlloyd/infinite-bach} and the JSB Chorales dataset\footnote{https://github.com/czhuang/JSB-Chorales-dataset}.
The JSB Chorales and Infinite Bach datasets contain MIDI files---382 for JSB Chorales and 498 for Infinite Bach---of chorales harmonized by J.S. Bach.
The MIDI files in Infinite Bach dataset are generally longer in duration allowing for approximately three times more samples overall compared to the JSB Chorales dataset.
Since the Gibbs sampling model is trained on JSB Chorales, we use the JSB Chorales dataset for benchmarking.

For both datasets, we preprocess the data by:
1) mapping MIDI to its piano roll representation using a sixteenth-note quantization, 
2) converting multi-track inputs into a single track by merging all tracks, and
3) splitting each MIDI into multiple 2 or 8 bar samples.

\subsection{Log Likelihood}

We calculate log likelihood using equation  \eqref{eq:mapping-joint}.
Since for each piano roll $x$ the inverse image $\pi^{-1}(x)$ is infinite, the sum cannot be calculated exactly; thus,
we calculate an approximate log likelihood for a subset of all possible edit sequences in $\pi^{-1}(x)$.
This value lower bounds the true log likelihood value. We compare this lower bound to the log likelihood for orderless NADE. Since our method removes notes as well, the proposed model is modeling a distribution with larger support so we do not expect the likelihood value of our method to be better than orderless NADE's. Our likelihood values show that---in the toy case when the sum can be sufficiently expanded---the likelihood lower bound value approaches that of orderless NADE.

Consider a graph where each vertex corresponds to a piano roll state and each edge corresponds to an edit event. 
A path in the graph corresponds to an edit sequence described in equation \eqref{eq:2}.  
As we traverse over a path, we calculate the log likelihood of the edit sequence corresponding to that path.

For each input $\mathcal{I}$ and target $\mathcal{T}$ pairing, we calculate our log likelihood over multiple levels, traversing over edit sequences of length $K+2d$ at level $d$.
$K$ is the minimum number of edit events needed to reach the target from the input.
All $K$ edit events are unique along time and pitch.
For level $d=0$, there exist $K!$ different edit sequences. 
We calculate the average log likelihood over a randomly chosen subset of these edit sequences and approximate it over all $K!$ edit sequences. 
During the traversal we keep track of the most probable (time, pitch) predictions that do not occur in the edit sequences, and add them to a pool $Q$. 
We keep these predictions as they will appear in the most probable edit sequence paths at level $d=1$.
For level $d=1$, we traverse down the same paths, but we add two edit events with the same time and pitch chosen from $Q$ to the path. 
This increases the path length to $K+2$ and results in the same target pianoroll since the two new edit events cancel out.
We approximate the log likelihood sum over all possible edit sequences. 
We repeat this for each (time, pitch) pair in $Q$.
This process can be repeated until level $d=D$ expanding our coverage of the edit sequence graph along the most probable paths.

\smallskip
We calculate the approximate log likelihood as:
\small
\begin{align*}
    \frac{1}{K}
    \log \sum_{d = 0}^{D}
    \sum_{Q}
    \frac{(K + 2d)!}{2^d}
    \frac{1}{\vert S \vert}
    \sum_{s \in S}
    p^{\operatorname{ES}}(s)
\end{align*}
\normalsize
where $S$ is a random subset of $K + 2d$ length edit sequences that can transform $\mathcal{I}$ to $\mathcal{T}$.\footnote{We divide the $K + 2d$ factorial by $2^d$ as we cannot ``remove" before we ``add" a note.}
As we increase the levels of our approximation, our log likelihood will converge towards orderless NADE which we see in Table \ref{tab:log-likelihood} at $d = 1$.

Since music completion is a task with high uncertainty, 
the large number of low probability predictions leads to underflow issues, which we avoid by using the log-sum-exp trick. Also, since log likelihood in this case is highly dependent on the number of notes in a piece, we compute an approximate \emph{notewise} log likelihood by dividing the approximate log likelihood by the minimum number of note additions and removals needed to reconstruct the target pianoroll.
We do not use log likelihood to compare our approach with Gibbs sampling used in Coconet as they use framewise log likelihood, which is different than our calculation \cite{huang2019counterpoint}.


\begin{table}[]
\centering
\resizebox{0.5\textwidth}{!}{%
\begin{tabular}{@{}lr@{}}
\toprule
Approach       & Notewise Approximate Log Likelihood \\ \midrule
ES-Net         & -0.635                                                   \\
Orderless NADE & -0.558                                                   \\ \bottomrule
\end{tabular}%
}
\caption{Notewise approximate log likelihood for reconstructing 10 missing notes from each test sample. 
}
\label{tab:log-likelihood}
\end{table}

\subsection{Quantitative Metrics}

We calculate several quantitative metrics to compare the quality of generated music using our approach, orderless NADE, and Gibbs sampling.
For each approach, we generate 3405 bars of music---the same number of bars in the training data---and compare them to the training data.
We generate the music by conditioning on 150 8-bar monophonic inputs.
We evaluate on the following metrics designed in \cite{dong2018musegan, dong2018pypianoroll}:
\begin{itemize}
    \itemsep0em 
    \item PC - Number of unique pitch classes used. Notes whole octaves apart from each other (e.g. C4 and C5) belong to the same pitch class.
    \item P - Number of unique pitches used.
    \item ISR - In-scale rate which is the proportion of all notes that lie in C Major\footnote{The C-Major scale was chosen arbitrarily.}.
    \item PR - Polyphonic rate which is the proportion of timesteps where the number of pitches being played is greater than or equal to 4.
\end{itemize}
We use pypianoroll \cite{dong2018pypianoroll} to calculate these values.

\begin{table}[h]
\centering
\resizebox{0.4\textwidth}{!}{%
\begin{tabular}{@{}lllll@{}}
\toprule
               & PC         & P           & ISR            & PR             \\ \midrule
Training Data  & 6          & 46          & 0.541          & 0.917          \\ \midrule
ES-Net         & \textbf{6} & \textbf{46} & \textbf{0.540} & \textbf{0.930} \\
Gibbs Sampling & \textbf{6} & \textbf{46} & 0.535          & 0.898          \\
Orderless NADE & 8          & 55          & 0.559          & 0.759          \\ \bottomrule
\end{tabular}%
}
\caption{Quantitative metrics for each approach. Closer to training data is better. Bold values are best between the three approaches.}
\label{tab:quant-metrics}
\end{table}

We observe that our approach and the Gibbs sampling approach both produce music that have similar characteristics to the dataset, while orderless NADE shows less similarity to the dataset.
As seen in Table \ref{tab:quant-metrics}, our approach  is the closest for all four metrics, with Gibbs sampling tying for number of unique pitches and pitch classes used.

\begin{table}[h!]
\centering
\resizebox{0.45\textwidth}{!}{%
\begin{tabular}{@{}ll|lll@{}}
\toprule
\multirow{2}{*}{} & \multirow{2}{*}{Bhattacharyya} & \multicolumn{3}{l}{Kolmogorov-Smirnov} \\ \cmidrule(l){3-5} 
                  &                                & df         & D           & p           \\ \midrule
ES-Net            & 0.028                          & 46         & 0.17        & 0.49        \\
Gibbs Sampling    & \textbf{0.021}                          & 46         & 0.13        & 0.83        \\
Orderless NADE    & 0.049                          & 46         & 0.17        & 0.49        \\ \bottomrule
\end{tabular}%
}
\caption{Various metrics for how far pitch appearance frequency is from the training data. Lower is better and bolded is best for Bhattacharyya distance. The Kolmogorov-Smirnov test is unable to show significant difference between any of the approaches and the training data.}
\label{tab:pitch}
\end{table}

\begin{figure}[t]
    \centering
    \includegraphics[width=0.9\linewidth]{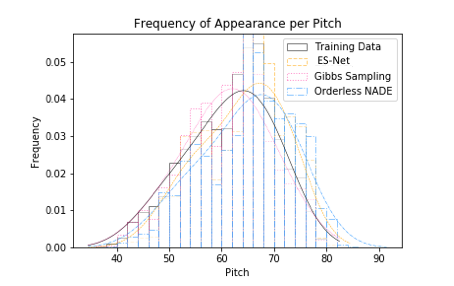}
    \caption[]{
    Frequency of occurrence for each pitch bin. Each bin is two pitches (i.e. one bin contains both pitch 31 and 32). 
    }
    \label{fig:ex4}%
\end{figure}

In Figure \ref{fig:ex4}, we plot the frequency of pitch values for each approach and compare with the distribution of pitches in the training data.
We observe that the distribution of pitches for all three approaches is very similar to that of the training data.
In Table \ref{tab:pitch}, we evaluate the similarity of each approach's pitch appearance frequency to the training data using various metrics. 
We calculate the Bhattacharyya distance \cite{bhattacharyya1943measure} showing Gibbs sampling as the closest to the training data and orderless NADE as the furthest from the training data.
We perform Kolmogorov-Smirnov tests and are unable to show significant differences between each approach and the training data.

\subsection{Human Evaluation}

\begin{figure*}[t!]
    \centering
    \subfigure[][]{
    \label{fig:improvement}
    \includegraphics[width=0.30\linewidth]{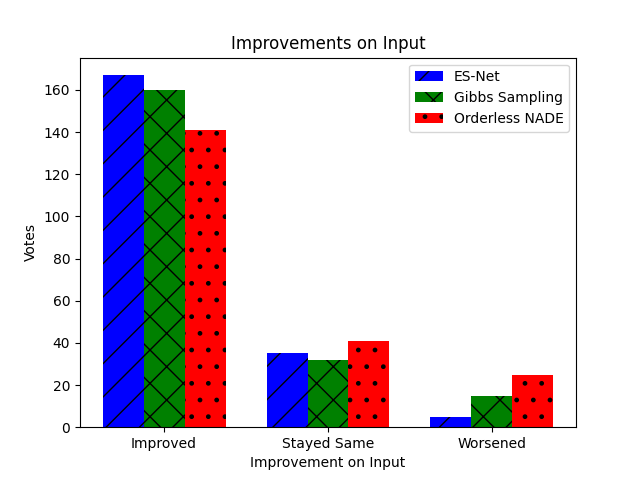}}
    \subfigure[][]{
    \label{fig:rank}
    \includegraphics[width=0.30\linewidth]{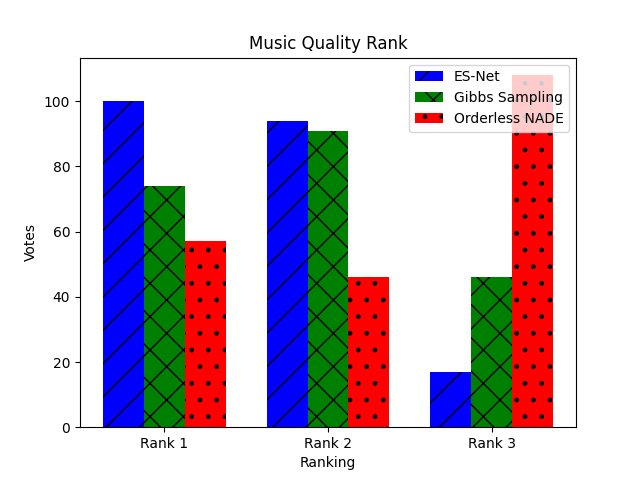}}
    \subfigure[][]{
    \label{fig:bach-likeness}
    \includegraphics[width=0.30\linewidth]{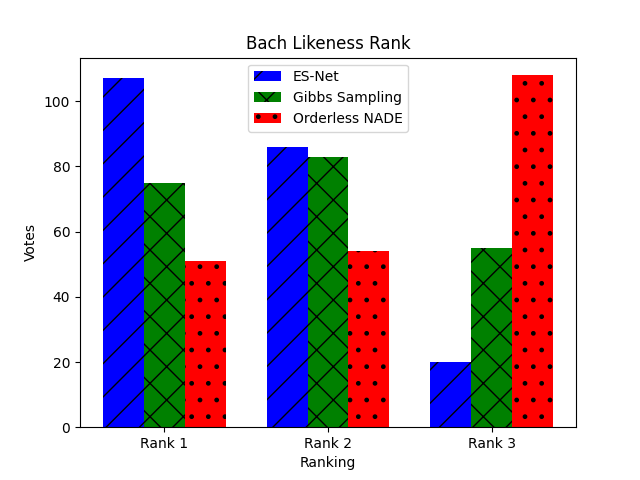}}
    \caption[A set of three subfigures.]{
    Human Survey Evaluation Ratings:
    \subref{fig:improvement} describes whether users thought a sample improved on the input.
    \subref{fig:rank} describes user rankings for music quality.
    \subref{fig:bach-likeness} describes user rankings for how similar a sample is to real Bach data. Bars are ordered from left-to-right as ES-Net, Gibbs Sampling, and Orderless NADE.
    }
    \label{fig:human}%
\end{figure*}

We conducted a human opinion test in order to compare our approach against orderless NADE and Gibbs sampling.
We generated 8 bar samples with a pitch range from 36 to 81.
We assume 4/4 time (i.e. 4 beats per bar) and quantize to 16 time steps per bar (i.e. 1/16th note).
We assume two notes continuous in time as one note.
For orderless NADE, we sample 400 times to generate samples that approximate to 4 pitches per time step.
Coconet optimizes the number of iterations it requires.
Since our approach allows for the model to both add and remove notes, there is no fixed number of iterations to run the model;
instead, the model eventually stabilizes and adds or removes the same set of notes repeatedly.
We sample for 10,000 iterations for our approach. When conducting the surveys, we chose a large number of iterations to ensure stabilization; through later experiments, however, we found that the output almost always stabilizes before 2000 iterations.

Each survey contained fifteen randomly chosen sets of comparisons where each set of comparisons contained a random sample from each of the three approaches.
Each of the samples in each set were randomly ordered.
All three samples in a set were conditioned on the same input track which was also given to the participant.
In order to simulate real user input, we created input tracks by taking two bar user inputs from the 
Bach Doodle Dataset \cite{bachdoodle2019}---a dataset of real user inputs to Coconet and its resulting composition---and 
repeated them four times to form eight bars.
Bach Doodle ranks their inputs based off of user feedback from the resulting Coconet composition; we chose an equal number of samples randomly from each feedback level.
Each survey contained the same fifteen sets of comparisons.
These inputs are monophonic (i.e. only one pitch per time step).
For each set of comparisons, users were asked
(a) if each sample improved on the input, 
(b) to rank the samples based on music quality, and
(c) to rank the samples based on similarity to music composed by Bach.


We receive a total of 207 ratings for question (a), 211 ratings for question (b), and 213 ratings for question (c).\footnote{Some users did not answer all three questions per set of samples. Partial or incomplete rankings were discarded.}
For question (a), we see that all approaches are comparable and each approach almost always improved the input as seen in Figure \ref{fig:improvement}.
For questions (b) and (c), we see in Figures \ref{fig:rank} and \ref{fig:bach-likeness} that our edit sequence approach is the best approach while the orderless NADE approach is the worst.
We perform a Kruskal-Wallis H-test across all ratings for questions (b) and (c).
We show that there is a statistically significant difference ($\mathcal{X}^{2}(2)=64.47$, $p < 0.001$ for question (a) and $\mathcal{X}^{2}(2)=73.07, p < 0.001$ for question (b)) between the three models.
We use the Wilcoxon signed-rank test to conduct a pairwise post-hoc analysis.
We show that there is a statistically significant difference ($p < 0.001$ for questions (b) and (c)) between our approach and both the Gibbs sampling and orderless NADE approaches.

\section{Future Work}\label{sec:future-work}
We currently trained on a limited number of datasets, both of which are based on Bach chorales.
There is no reason, however, that our approach should be limited to any feature of Bach.
By training on other datasets, we will be able to evaluate how well our approach generalizes.

We show that allowing the model to remove notes increases music quality
which we believe is due to the model correcting its past mistakes.
During our training process, we generate random notes in order to mimic those mistakes.
Rather than merely mimicking those mistakes, however, we can generate real mistakes by feeding outputs from the model back into itself.
We believe that this self-adversarial training paradigm will allow the model to capture more realistic sampling mistakes and further improve performance.

Our current data representation does not convey features such as note velocity, repeated notes, or explicit note duration.
These features, however, can add to the technical and emotive quality of music.
We can map these new features as additional channels and concatenate this information with our existing piano roll.
This new data representation will allow our model to learn from these new feature dimensions and produce more expressive and technically challenging music.

An advantage of our algorithm is the ease with which we can extend our approach to other use cases.
For instance, currently our model generates fixed length outputs depending on the length of the training samples.
In this way, we can extend user melodies up to a fixed length;
however, we never explicitly train our model to extend inputs.
By augmenting our dataset so that the latter portion of each sample is masked out, we can explicitly train our model to extend melodies.
Then, during sampling, we can generate a fixed length output, feed the latter portion of that output back into the model to generate a new output, concatenate those two outputs together, and repeat.
This would allow us to extend melody repeatedly rather than up to a fixed length output.

\section{Conclusion}\label{sec:conclusion}

We show that by modeling removal of notes, we can train a model to produce better music by fixing past mistakes and preventing accumulation of errors.
We discuss how our note-by-note approach allows for a finer degree of control and better human and AI collaboration.
We demonstrate how to map an edit sequence representation into a piano roll representation and how we can use that to model a distribution of musical pieces.
We discuss how we train our model by masking and adding erroneous notes and how we sample from our model during inference.
Finally, we show through quantitative metrics and human evaluation that our approach is able to generate musical compositions that are of better quality than orderless NADE and Gibbs sampling.

\bibliography{bibliography}

\end{document}